\documentclass{moriond}

\usepackage{xspace}
\usepackage{amsmath}
\usepackage{caption}
\usepackage{lineno}
\usepackage{hyperref}

\def\be{\begin{equation}}
\def\ee{\end{equation}}
\def\bea{\begin{eqnarray}}
\def\eea{\end{eqnarray}}

\newcommand{\Photo}{\includegraphics[height=35mm]{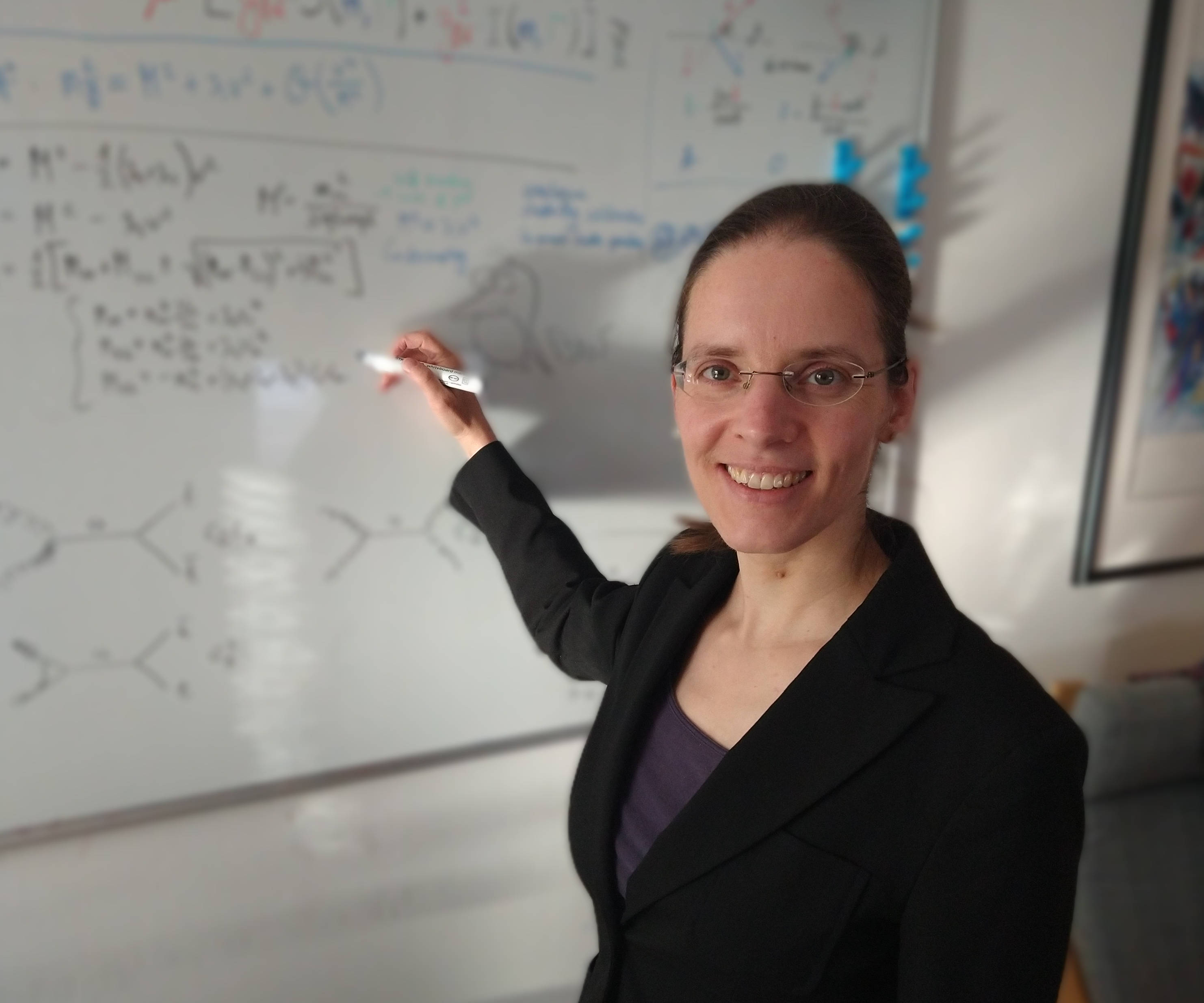}}
\newcommand{\toytoponium}{\ensuremath{\eta_t}\xspace}
\newcommand{\toponium}{\ensuremath{t\bar{t}_{\text{GFRW}}}\xspace}
\newcommand*{\ttbar}{\ensuremath{t\bar{t}}\xspace}
\newcommand*{\mttbar}{\ensuremath{m_{t\bar{t}}}\xspace}
\newcommand{\chel}{\ensuremath{c_\mathrm{hel}}\xspace}
\newcommand{\chan}{\ensuremath{c_\mathrm{han}}\xspace}
\newcommand{\met}{\ensuremath{E_\mathrm{T}^\mathrm{miss}}\xspace}

\begin{document}
\vspace*{4cm}
\title{Observation of a cross-section enhancement near the $t\bar{t}$ production threshold\\
in $\sqrt{s}=13$~TeV $pp$ collisions with the ATLAS detector}

\author{Janna Katharina Behr\\ on behalf of the ATLAS Collaboration\footnote{Copyright 2026 CERN for the benefit of the ATLAS Collaboration. CC-BY-4.0 license.}}

\address{Deutsches Elektronen-Synchrotron DESY,
Notkestr. 85, 22607 Hamburg, Germany}

\maketitle\abstracts{
A significant excess of \ttbar events near the production threshold was observed in LHC Run-2 data by the ATLAS Collaboration. It is consistent with the formation of \ttbar quasi-bound states, which were first hypothesised almost 40 years ago. This contribution summarises the experimental results and outlines a path toward further characterisation of the excess.}

The top quark stands out among the fermions of the Standard Model due to its large mass and consequently extremely short lifetime of approximately $5 \times 10^{-25}$ seconds. This lifetime is shorter than the average time required for the formation of hadronic bound states ($3 \times 10^{-24}$~s) and spin decorrelation ($3 \times 10^{-21}$~s).
This means that the top quark does not form hadronic bound states and its spin state can be accessed from the angular distributions of its decay products.
If a top-antitop quark pair (\ttbar) is produced with an invariant mass \mttbar close to the kinematic threshold $2 m_{\mathrm{top}}$, the top and antitop quark typically have small relative velocities and can be described by non-relativistic QCD (NRQCD). If the \ttbar system is in a colour-singlet state, the NRQCD Coulomb potential is attractive and the \ttbar system can form short-lived quasi-bound states, colloquially dubbed ``toponia''. This state does not decay via quark-antiquark annihilation but via the weak decay of one of the constituent quarks. It manifests itself as a narrow enhancement of the \ttbar cross-section below the production threshold, dominated by pseudo-scalar $^1S_0$ states. These and other NRQCD modifications of the \ttbar threshold were predicted as early as 1987~\cite{Fadin:1987wz} but assumed to be impossible to observe at the Large Hadron Collider (LHC). They were hence not fully included in standard Monte Carlo (MC) models of \ttbar production at the LHC.
Recently, the ATLAS~\cite{ATLAS:2026dbe} and CMS~\cite{CMS:2025kzt} Collaborations reported the observation of significant excesses of \ttbar events near the production threshold, consistent with models of \ttbar quasi-bound-state formation.

\section{Data and simulated samples}

The measurement is based on proton–proton ($pp$) collision data recorded with the ATLAS detector~\cite{ATLAS:2008xda} at a centre-of-mass energy $\sqrt{s}= 13$~TeV,
with an integrated luminosity of 140 fb$^{-1}$.

\subsection{Baseline MC models of \ttbar production}

In the baseline MC model, \ttbar production is modelled at next-to-leading order (NLO) in perturbative QCD (pQCD). No NRQCD effects, besides those entering at NLO in pQCD, are considered in this model.
Events are generated using the \texttt{hvq} model in \textsc{PowhegBox v2}, interfaced with \textsc{Pythia8.2}.
The sample is normalised to the inclusive cross-section of $834^{+37}_{-43}$~pb, obtained at NNLO+NNLL precision.
The differential predictions for the kinematic variables of the top-quarks in this and related systematic variation samples are improved by correcting them to more accurate differential predictions calculated at NNLO-QCD+NLO-EW accuracy with \texttt{MATRIX} and \texttt{HATHOR}, respectively. The corrections are applied via a two-dimensional reweighting in \mttbar and $\cos\theta^*$, where $\theta^*$ is the angle between the momentum of the top quark in the \ttbar centre-of-mass frame and the momentum of the reconstructed \ttbar system in the laboratory frame.

An alternative pQCD \ttbar sample based on the \texttt{bb4l} model in \texttt{PowhegBoxRes} is used to estimate the impact of off-shell top-quark decays. This model includes off-shell
and non-resonant contributions as well as exact spin correlations at NLO accuracy and simulates the
inclusive production of $b^+b^-\ell^+\ell^-\nu^+\nu^-$ final states.

\subsection{Extended MC model of \ttbar production}

The extended MC model includes NRQCD effects in addition to the baseline MC model of pQCD \ttbar production.
NRQCD effects at the threshold are modelled following the approach in Ref.~\cite{Fuks:2024yjj}.
In this framework, the formation of colour-singlet states is described using the NRQCD Green’s function in the Coulomb gauge. The ratio of this Green's function to the one for free top quarks is used to reweight the LO matrix elements obtained with \textsc{MadGraph}. Only $^{1}S_0$ \ttbar production from gluon-gluon initial states is considered.
The Green’s function reweighting is applied only
to events with $\mttbar < 350$~GeV and a top-quark momentum magnitude $p^*$ in the \ttbar rest frame $<50$~GeV to minimise the overlap with the pQCD \ttbar samples.
This sample, which is referred to as \toponium, is normalised to a cross-section of 6.43 pb from analytical calculations~\cite{Fuks:2021xje}.
%
%

\section{Analysis strategy}

The measurement is conducted in final states with two oppositely charged leptons (electrons and/or muons) and at least two jets.
Candidate events are required to contain exactly two oppositely charged leptons with transverse momentum $p_T > 10$~GeV, with at least one lepton satisfying $p_T > 28$~GeV, and at least two jets with $p_T > 25$~GeV. At least one of the jets must be $b$-tagged, i.e. identified as originating from a $b$-quark. Additionally, in events with same-flavour leptons, the dilepton invariant mass must be $>15$~GeV, not within the range $81 - 101$~GeV around the $Z$-boson mass, and the event is required to have $\met > 60$ GeV to reduce the background from $Z$+jets events. Only events passing these requirements are considered for the signal regions (SRs) of the measurement.
These contain small contaminations from non-\ttbar processes, with 4\% of events in these regions arising from $tW$ production, 1.5\% from events with objects falsely identified as leptons (``Fakes''), and 0.8\% from $Z$+jets production. 

The four-vectors of the top and antitop quarks are reconstructed by first selecting the two $b$-jet candidates from their decays. If more than two of the jets in the event are $b$-tagged, the two highest-$p_T$ $b$-tagged jets are selected. If there is only one $b$-tagged jet, the highest-$p_T$ jet among the remaining untagged ones is selected.
Next, the four-momenta of the two neutrinos from the leptonically decaying $W$-bosons are obtained from the selected $b$-candidate jets, charged leptons, and the total \met in the event by applying constraints from the $W$ boson and top quark masses in an analytical approach known as the Ellipse Method~\cite{Betchart:2013nba}.
The resolution of the reconstructed \mttbar is about 22\% at the \ttbar threshold and improves to 18\% around 500~GeV.

Only events with $\mttbar<500$~GeV are considered for the measurement. The events passing all selection requirements are categorised into nine SRs based on two angular observables, \chel and \chan, which are sensitive to \ttbar spin correlations. The variable \chel is the scalar product of the unit vectors of the momenta of the two leptons after Lorentz boosting the respective four-momenta first into the \ttbar centre-of-mass frame, and then separately into their parent (anti)top-quark rest frames~\cite{ATLAS:2023fsd}. The observable \chan is the cosine of the same angle, where the sign of the
component of the lepton momentum along the top-quark flight direction is flipped~\cite{Aguilar-Saavedra:2022uye}.
These observables serve to discriminate spin-singlet \ttbar states from pQCD \ttbar events.

\section{Results}

\begin{figure}
 \includegraphics[width=0.95\linewidth]{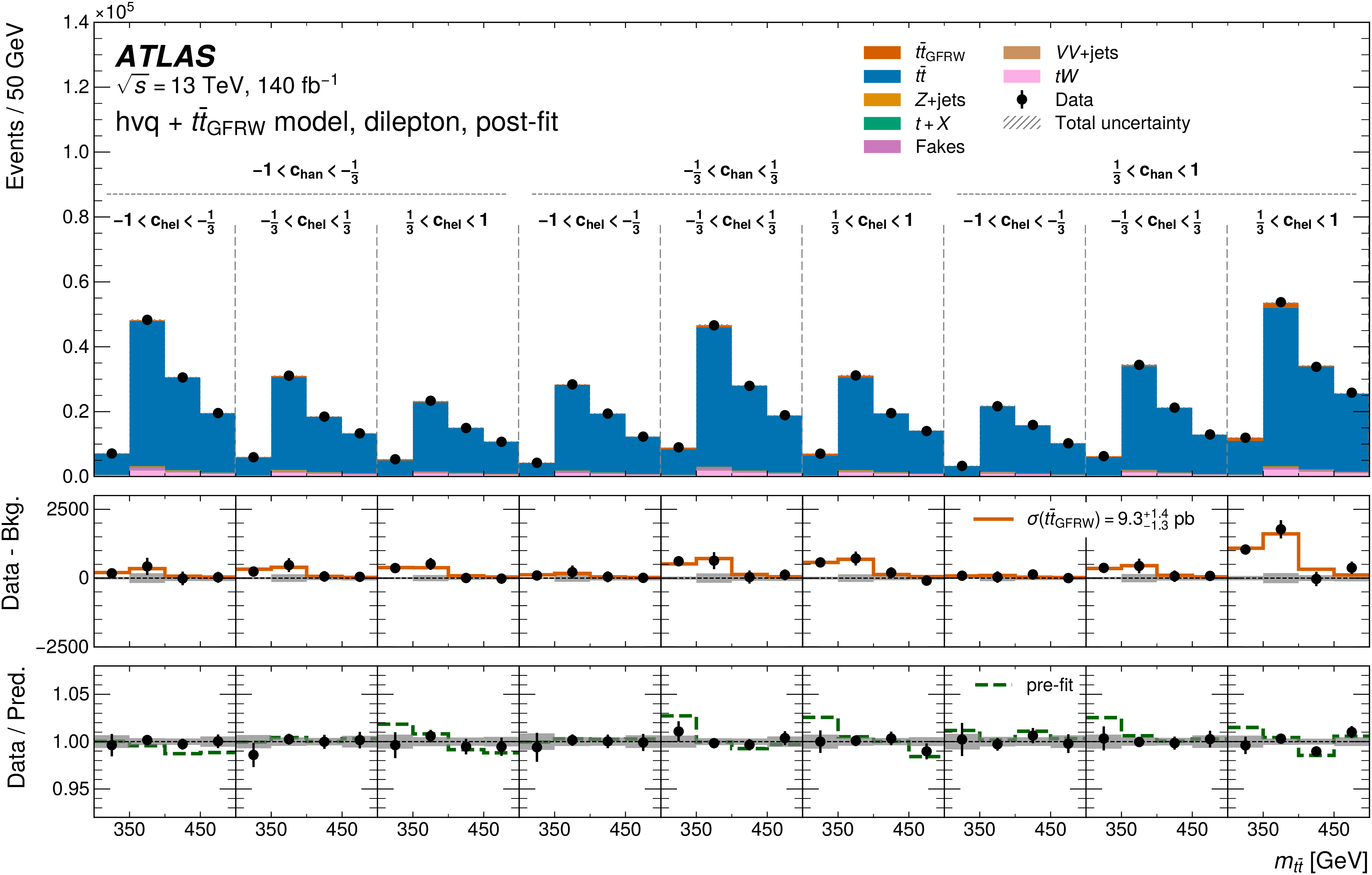}
  \caption[]{Observed (points with statistical error bars) and expected (stacked coloured histograms) \mttbar distributions after the fit to data in the SRs (upper panel).
  The middle panel contains a comparison between the predicted \toponium distributions and the data, from which the baseline pQCD \ttbar contribution and background processes were subtracted.
  The bottom panel shows the ratio of the data and the extended \ttbar model, which includes the \toponium distribution.
  The grey hashed and shaded bands represent the total systematic uncertainty in the prediction. The dashed line represents the data/MC ratio before the fit.~\cite{ATLAS:2026dbe}}
  \label{fig:postfit}
\end{figure}

The agreement between the data and the predictions of the baseline and extended models, respectively, is quantified using a binned profile-likelihood fit of the \mttbar distributions in the nine SRs (Figure~\ref{fig:postfit}).
The normalisation of both the pQCD \ttbar and \toponium contributions are free-floating in the fit.
The baseline model without quasi-bound-state
contributions is rejected with an observed (expected) significance of over $8\sigma$ ($6\sigma$).
The cross-section for the \toponium contribution extracted from its normalisation factor in the fit with the extended model is
\be
\sigma(\toponium) = \; 9.3^{+1.4}_{-1.3}~\mathrm{pb} \; = \; 9.3^{+1.1}_{-1.0}~\mathrm{(stat.)} \pm 0.8~\mathrm{(syst.)}~\mathrm{pb}. 
\ee
This value is $(45^{+21}_{-20})\%$ larger than the calculated value of 6.43 pb, in line with a slight pre-fit excess of the data compared to the extended model at low \mttbar.

The systematic uncertainty in the fitted cross-section is dominated by modelling uncertainties in the pQCD \ttbar and \toponium components of the extended model. For the \toponium component, the dominant effects are the uncertainties in the modelling of final- and initial-state radiation, respectively, which are largest in the lowest \mttbar bins and decrease towards higher \mttbar. For pQCD \ttbar production, the largest uncertainty in terms of its impact on the measured cross-section is that related to the scale choice in the NNLO QCD reweighting. Only moderate shifts and constraints are observed for the nuisance parameters related to the dominant modelling and experimental uncertainties in the profile-likelihood fit.

An alternative set of results has been obtained using \texttt{bb4l} for the pQCD \ttbar prediction.
The alternative baseline model is rejected at a significance of more than $8\sigma$ ($7\sigma$ expected).
The \toponium cross-section is measured to be $8.5^{+1.2}_{-1.1}$~pb, consistent with the cross-section obtained using \texttt{hvq} for the baseline model.

For a more direct comparison with Ref.~\cite{CMS:2025kzt}, another set of results is obtained using a simplified model, in which NRQCD effects are approximated via a pseudoscalar resonance \toytoponium.
Using this model along with the \texttt{hvq} baseline, a cross-section $\sigma(\toytoponium) = 13.1^{+1.9}_{-1.7}$~pb is obtained. The larger observed cross-section compared with the \toponium model can be explained by the differences between the \mttbar distributions in the SRs, which are understood to be partially related to the fact that no upper bound is applied on the \mttbar at the parton level for the simplified model.

\section{Outlook}

Further efforts toward a more complete MC model of NRQCD effects near the \ttbar threshold are needed to reconcile the results obtained with the \toytoponium model and the \toponium model. Future models should also include $P$-wave and colour-octet, along with higher-order NRQCD contributions.
Improved models of the pQCD \ttbar component, which include NNLO QCD corrections in addition to an accurate treatment of off-shell top-quark decays, are needed.
Importantly, an improved matching between NRQCD and pQCD predictions is essential to further characterise the nature of the excess in future precision measurements by the ATLAS and CMS Collaborations using data from LHC Run 3.

\section*{Acknowledgments}

I acknowledge support by the Helmholtz Young Investigators Group VH-NG-1503 and by the Deutsche Forschungsgemeinschaft (DFG, German Research Foundation) under Germany's Excellence Strategy EXC 2121 Quantum Universe 390833306.

\section*{References}
\bibliography{KatharinaBehr}


\end{document}